\documentclass[%
article
superscriptaddress,
nofootinbib,
 amsmath,amssymb,
 aps,
prd,
floatfix,
]{revtex4-2}

\newcommand{\CS}{\text{CS}}

\newcommand{\lat}{\text{lat}}
\newcommand{\gap}{\text{gap}}

\newcommand{\I}{\mathrm{i}}
\newcommand{\Tr}{\mathrm{Tr}}
\newcommand{\To}{\rightarrow}
\newcommand{\lc}{\lim_{\chi}}

\newcommand{\sumcorrmin}{\sum_{
			I_1...I_{2k}}}

\newcommand{\acorrkmin}{\Gamma_{\begin{matrix}
			x_1...x_{2k}\\
			I_1...I_{2k}
\end{matrix}}}
\newcommand{\acorrkbarmin}{\bar{\Gamma}_{\begin{matrix}
			x_1...x_{2k}\\
			I_1...I_{2k}
\end{matrix}}}

\newcommand{\cond}[2]{\begin{footnotesize}\begin{matrix}#1 = 0\\#2 = 0\end{matrix}\end{footnotesize}}
\usepackage{tikz}
\usetikzlibrary{decorations.pathreplacing}
 \usetikzlibrary{plotmarks}
\usepackage{subfigure}
\usepackage{booktabs}
\usepackage{mathtools}
\usepackage{dsfont}
\usepackage{xcolor}

\usepackage{graphicx}
\usepackage{dcolumn}
\usepackage{bm}

\usepackage{braket}
\usepackage{multirow}
\usepackage{mathtools}
\usepackage{xcolor}
\usepackage[normalem]{ulem}

\begin{document}

\title{Gap in the Dirac spectrum and quark propagator symmetries in lattice QCD}

\author{Marco Catillo}
\email{mcatillo@phys.ethz.ch}

\affiliation{Institute for Theoretical Physics, ETH Z\"urich, 8093 Z\"urich, Switzerland}
 
\date{\today}

\begin{abstract}
Recent studies on lattice QCD have shown the emergence of large symmetries at high temperature. 
This includes not only the restoration $SU(n_F)_L \times SU(n_F)_R$, but also the effective emergence of an  unexpected symmetry group, namely $SU(2)_{CS}$, which contains $U(1)_A$ as subgroup. 
At the same time, at high $T$, a gap in Dirac spectrum appears. 
As it is argued in several works of \textit{L. Glozman et al.}, there should be a connection between a gap in the Dirac spectrum and the presence of $SU(2)_{CS}$.
In this paper, we analyze whether the quark propagator can be invariant under  $SU(n_F)_L \times SU(n_F)_R$ and $SU(2)_{CS}$ transformations, in case of a gap in the Dirac spectrum, and consequently the invariance of hadron correlators, giving the condition for a quark propagator to be invariant under $SU(2)_{CS}$. 
\end{abstract}
\maketitle



\section{Introduction}\label{sec:intro}

As it is well-known, Chiral and axial symmetry are symmetries of the QCD Lagrangian. 
However, in QCD chiral symmetry is broken spontaneously at low temperature, i.e. $T<T_c$ and axial symmetry is violated by the anomaly. 
Nevertheless, in lattice studies is evident that chiral symmetry get restored at high temperature and in many works (see e.g. \cite{Bazavov:2012qja,Tomiya:2016jwr}) also the axial symmetry seems to emerge at high temperature (however we also refer to \cite{Ding:2020xlj} for a recent study on this issue). 
However, this is not the whole story. 
In a range of temperatures, approximately $T_c - 3\,T_c$, also a further larger symmetry appears and this is what has been found in \cite{Rohrhofer:2017grg,Rohrhofer:2019qal,Rohrhofer:2019qwq}. 
This symmetry includes the $U(1)_A$ and it has been called $SU(2)_{CS}$, see \cite{Denissenya:2014poa,Denissenya:2015mqa,Denissenya:2015woa,Rohrhofer:2017grg,Rohrhofer:2019qal,Rohrhofer:2019qwq,Glozman:2015qva}. 
However, the full $SU(2)_{CS}$ is not a symmetry of QCD Lagrangian, but still emerges at high temperature in the calculation of hadron correlators. 

The important feature of QCD at high temperature is that, the lowest eigenmodes as well as zero modes of the Dirac operator become naturally suppressed. 
This can explain why an effective emergence of $U(1)_A$ appears at high temperature and, from the Banks-Casher relation \cite{Banks:1979yr}, why $SU(n_F)_L \times SU(n_F)_R$ is restored. 

Regarding $SU(2)_{CS}$, this fact is more fascinating, since not only it becomes evident at high temperature \cite{Rohrhofer:2017grg,Rohrhofer:2019qal,Rohrhofer:2019qwq}, when the lowest eigenmodes are suppressed, but its emergence is very explicit when the lowest eigenmodes are removed manually from the quark propagator in the calculation of several hadron correlators, see \cite{Denissenya:2014poa,Denissenya:2015mqa,Denissenya:2015woa}.

In this paper, 
we generalize the results of \cite{Lang:2018vuu,Catillo:2019jrl} and
we consider precisely the case where there is a gap in the Dirac spectrum. 
For simplicity, we will not consider possible anomaly terms in the action, which will arise the $U(1)_A$ breaking. 
In this situation, we study the symmetries of the quark propagator starting from its formulation on the lattice. 
The reason is that, if the quark propagator has a symmetry, then also observables, that can be written as only function of it (e.g. hadron correlators), contain such symmetry. 
We will show that, when there is a gap in the Dirac spectrum, the quark propagator becomes $SU(n_F)_L \times SU(n_F)_R$ and $U(1)_A$ invariant. 

Driving by this fact, we observe, upon a gap in the Dirac spectrum, 
which conditions the quark propagator has to satisfy if we want it $SU(2)_{CS}$ invariant and having the emergence of $SU(2)_{CS}$.


\section{Some preliminaries}\label{sec:prelim}

We consider the euclidean formulation of QCD on the lattice with $n_F$ degenerate quark flavors.
However we will not examine the case where interaction terms among quark with different flavors are present in the action (e.g. the presence of a 't Hooft term) and we will not consider the presence of zero modes in the theory. 
In this case, the fermionic action can be split as $S_F = \sum_{i=1}^{n_F} S_i$, where $S_i$ is the action of a single quark flavor, namely $S_i = a^4\sum_{x,y} \bar{\psi}(x) D_{\lat}^{(m)}(x,y)\psi(y)$, where $a$ is the lattice spacing and $m$ is the mass of a single quark flavor. 
The full action is, therefore, given by $S = S_G + \sum_{i=1}^{n_F} S_i$, where $S_G$ is the gauge field action.\\
Moreover we call $D_{\lat}^{(m)}$ the Dirac operator on the lattice for a given quark flavor with mass $m$. 
Taking, for example, the Wilson discretization of such operator, it satisfies the relation $D_{\lat}^{(m)} = \omega D_{\lat} + m$, with $\omega = 1 - am/2$ and $D_{\lat}$ is the massless Dirac operator (see \cite{Chandrasekharan:1998wg,Gattringer:2010zz}). 
$D_{\lat}$ also satisfies the Ginsparg-Wilson equation and the $\gamma_5$-hermiticity, i.e. 

\begin{equation}
\{D_{\lat},\gamma_5\} = a\, D_{\lat} \gamma_5 D_{\lat},\quad
D_{\lat}^{\dagger} = \gamma_5 D_{\lat}\gamma_5.
\label{eq:gw}
\end{equation}

\noindent
The $\gamma_5$-hermiticity implies that the non-zero eigenvalues of $D_{\lat}$ come in pairs $(\lambda_n,\lambda_n^{*})$, since

\begin{equation}
D_{\lat} v^{(+)}_n = \lambda_n v^{(+)}_n,\quad
D_{\lat} v^{(-)}_n = \lambda_n^{*} v^{(-)}_n,
\label{eq:eig1}
\end{equation}

\noindent
where $v^{(-)}_n = \gamma_5 v^{(+)}_n$ and we have the following normalization, $(v^{(\pm)}_n,v^{(\pm)}_m) = (1/V)\delta_{nm}$ and $(v^{(\mp)}_n,v^{(\pm)}_m) = 0$, with $V$ the total number of eigenvalues. 
Instead the Ginsparg-Wilson equation (together with the $\gamma_5$-hermiticity) tells us that the eigenvalues lie on a circle with equation: $\lambda_n + \lambda_n^* = a\vert \lambda_n\vert^2$. 
Denoting with $\eta_n = \vert\lambda_n\vert$, then
such eigenvalues can be written as
$\lambda_n = a\eta_n^2/2 + \I\eta_n\sqrt{1-(a\eta_n/2)^2}$ for $\eta_n \in (0,2/a]$. 

The inverse of $D_{\lat}^{(m)}$ can be written as

\begin{equation}
D_{\lat}^{(m)\,-1} = \sum_n \left[(\omega\lambda_n + m)^{-1}v^{(+)}_n v^{(+)\,\dagger}_n +\right. \left.(\omega\lambda_n^* + m)^{-1}v^{(-)}_n v^{(-)\,\dagger}_n\right].
\label{eq:qprop1}
\end{equation}

\noindent
$D_{\lat}^{(m)\,-1}$ is basically the quark propagator and it is the main quantity for our interest. 
The reason is that, $D_{\lat}^{(m)\,-1}$ is used to compute hadron correlators and it is also equal to the mean $\left\langle \psi_i \bar{\psi}_i\right\rangle_F = D_{\lat}^{(m)\,-1}$, for a given quark flavor $i$. 
The subscript $F$ stands for the average over the fermionic fields, which is weighted by the factor $\exp(-S_F)$.
The quantity $\left\langle \psi_i \bar{\psi}_i\right\rangle_F$ is connected with the chiral condensate $\Sigma$, which is the order parameter for the chiral symmetry breaking.
The thing that we want to show is how $D_{\lat}^{(m)\,-1}$ transforms under chiral, axial and $SU(2)_{CS}$ transformations and which are the criteria for having the emergence of such symmetries in observables and hadron correlators. 
For doing so, we need to rewrite $D_{\lat}^{(m)\,-1}$ in a more suitable way, separating different kind of contributions. 
At first we see that, using the expression of 
$\lambda_n$ in terms of $\eta_n$,
we can write the coefficients in eq. (\ref{eq:qprop1}) as, $(\omega\lambda_n + m)^{-1} = g(m,\eta_n)+h^r (m,\eta_n) + \I h^i(m,\eta_n)$,
where 

\begin{equation}
\begin{split}
&g(m,\eta) = \frac{m}{k^2\eta^2 + m^2},\quad
h^r (m,\eta) = \frac{1}{2}\frac{\omega\,a\,\eta^2}{k^2\eta^2 + m^2},\quad h^i (m,\eta) = \frac{\omega\,\eta}{k^2\eta^2 + m^2}\sqrt{1-\left(\frac{a\,\eta}{2}\right)^2},
\end{split}
\label{eq:ghh}
\end{equation}

\noindent
with $k = 1 - (am/2)^2$ and use that $(\omega\lambda_n^* + m)^{-1} = (\omega\lambda_n + m)^{-1\,*}$. 
The functions in eq. (\ref{eq:ghh}) are basically the same introduced in \cite{Lang:2018vuu,Catillo:2019jrl}, generalized on the lattice, but with the addition of a further function, $h^r (m,\eta)$, which comes from the lattice discretization. 

Now we can decompose $D_{\lat}^{(m)\,-1}$ as 

\begin{equation}
D_{\lat}^{(m)\,-1} = \left. D_{\lat}^{(m)\,-1}\right\vert_{\cond{h^r}{h^i}} +
\left.D_{\lat}^{(m)\,-1}\right\vert_{\cond{g}{h^i}} +
\left.D_{\lat}^{(m)\,-1}\right\vert_{\cond{g}{h^r}},
\label{eq:qprop2}
\end{equation}

\noindent
where

\begin{equation}
\begin{split}
&\left.D_{\lat}^{(m)\,-1}\right\vert_{\cond{h^r}{h^i}} = \sum_n g(m,\eta_n)\left[
v^{(+)}_n v^{(+)\,\dagger}_n + v^{(-)}_n v^{(-)\,\dagger}_n
\right],\\
&\left.D_{\lat}^{(m)\,-1}\right\vert_{\cond{g}{h^i}} = \sum_n h^r(m,\eta_n)\left[
v^{(+)}_n v^{(+)\,\dagger}_n + v^{(-)}_n v^{(-)\,\dagger}_n
\right],\\
&\left.D_{\lat}^{(m)\,-1}\right\vert_{\cond{g}{h^r}} = \sum_n \I h^i(m,\eta_n)\left[
v^{(+)}_n v^{(+)\,\dagger}_n - v^{(-)}_n v^{(-)\,\dagger}_n
\right].\\
\end{split}
\label{eq:qprop3}
\end{equation}

\noindent
The three functions, $g(m.\eta)$, $h^r(m.\eta)$ and $h^i(m.\eta)$, have different behavior for small and large values of $\eta$. 
Therefore they are important for us to analyze which part of the quark propagator $D_{\lat}^{(m)\,-1}$ is dominant through different distributions of the Dirac eigenvalues. 
Naively speaking, we can observe that for $\eta \rightarrow \infty$, we have that $g(m,\eta)\sim 1/\eta^2$, $h^r(m,\eta)\sim \omega a/2k^2$ and $h^i(m,\eta)\sim 1/
\eta$. 
Hence $g(m,\eta)$ tends to be more suppressed for large $\eta$, with respect $h^r(m,\eta)$ and $h^i(m,\eta)$. 
Vice versa for small $\eta$, then $h^r(m,\eta)\sim 0 $ and $h^i(m,\eta)\sim 0 $, while $g(m,\eta)$ blows up. 
Furthermore, if we take the continuum limit ($a\rightarrow 0 $) and afterwards the massless limit ($m\rightarrow 0$), then $g(m,\eta)$ becomes a $\delta$-function centered in $\eta = 0$. 
Therefore $g(m,\eta)$ tends to select the smallest eigenvalues. 
This point is crucial for the Banks-Casher relation \cite{Banks:1979yr} and the chiral symmetry breaking, as we will clear in the next section.

\section{Gap in the Dirac spectrum}\label{sec:gap}

In this section, we want to show, from the Bank Casher relation \cite{Banks:1979yr}, how in presence of a gap in the Dirac spectrum, the only relevant contribution in the quark propagator, in eq. (\ref{eq:qprop2}), is given by $\left.D_{\lat}^{(m)\,-1}\right\vert_{\cond{g}{h^r}}$, while the others, in eq. (\ref{eq:qprop3}), disappear in the following limit order
$
\lc\equiv \lim_{m\To 0}\lim_{a\To 0}\lim_{V\To\infty},
$ 
where, for convenience, we define $\lc$, as the chiral limit.

Considering a gauge configuration $\{U_{\mu}(x)\}$, which is a set of gauge links in every point of the lattice, we can evaluate then the Dirac operator $D_{\lat}^{(m)}$
 as well as the quark propagator $D_{\lat}^{(m)\,-1}$.
From the Dirac operator obtained from such gauge configuration, we can get the distribution of the variable $\eta$ as

\begin{equation}
\rho_a(m,V,\eta) = \frac{2}{V}\sum_{n:\eta_n \neq 2/a} \delta(\eta - \eta_n),
\label{eq:distrrho1}
\end{equation}

\noindent
where we excluded the eigenvalues $\eta_n = 2/a$, because 
these eigenvalues give some divergence problems in the continuum limit, as pointed out in \cite{Gattringer:2010zz}. 
$\rho_a(m,V,\eta)$ is positive for all $\eta, m, V$ and $a$, and its normalization is given by $\int_0^{\infty}d\eta\,\rho_a(m,V,\eta) = 1 - \varepsilon$\footnote{$\varepsilon = (2l/V)$, where $l$ is the number of eigenvalues $2/a$ in the spectrum. By definition $0\leq l\leq (V/2)$, hence $1-\varepsilon\leq 1$.}. 
We need also to point out that all $\eta_n$ are strictly positive in the sum in eq. (\ref{eq:distrrho1}), which justifies the normalization in front of the equation, since there are exactly $V/2$ values of $\eta_n$. 

Using eq. (\ref{eq:distrrho1}), we can write the Banks-Casher relation as $
\lc \langle \Sigma(m,a,V)\rangle_G = (\pi/2)\langle \rho_0(0,0)\rangle_G$,
where we have indicated for convenience $\rho_a(m,\eta) = \lim_{V\To\infty} \rho_a(m,V,\eta)$, and the factor $1/2$ comes from the normalization given for $\rho_a(m,V,\eta)$ in eq. (\ref{eq:distrrho1}). The average $\langle\cdot\rangle_G$ is over the gauge fields and weighted by the factor $\exp(-S_G)$. 
$\Sigma(m,a,V)$ is instead the chiral condensate on the lattice, which is given by \cite{Gattringer:2010zz}

\begin{equation}
\Sigma(m,a,V) = -\sum_{x}\langle \bar{\psi}_i\left(\mathds{1} - \frac{a}{2}D_{\lat}\right)\psi_i\rangle_F= \frac{1}{\omega}\Tr\left(D_{\lat}^{(m)\,-1} - \frac{a}{2V}\mathds{1}\right),
\label{eq:chiralcond1}
\end{equation}

\noindent
where $i$ indicates a given flavor (no sum is understood).
In the second line of eq. (\ref{eq:chiralcond1}), we are basically removing by hand the contribution of $2/a$ eigenvalues from the trace. 
We can, in principle, define a Banks-Casher relation which is also valid for a given gauge configuration, in the sense that $\lc\Sigma(m,a,V) =(\pi/2)\rho_0(0,0) $, where the meaning of this limit will be clear in a while. 
In order to see this point in detail, we can use the result (\ref{eq:qtrace2}) of Appendix \ref{app:trace} and rewrite $\Sigma(m,a,V)$ as 

\begin{equation}
\begin{split}
&\Sigma(m,a,V) = \frac{1}{\omega}\left(\Tr (D_{\lat}^{(m)\,-1}) - \frac{a}{2}\right)=
\frac{2}{\omega}\int_0^{\infty}d\eta\,\rho_a(m,V,\eta)(g(m,\eta) + h^r(m,\eta))+a\frac{l}{\omega V}-\frac{a}{2\omega},
\end{split}
\label{eq:chiralcond2}
\end{equation}

\noindent
where, we observe that $h^i (m,\eta)$ does not contribute and the further terms proportional to $a$ will vanish in the limits $\lim_{a\To 0}\lim_{V\To\infty}$.
Hence, from eq. (\ref{eq:qtrace1}), we get

\begin{equation}
\Sigma(m,a,V) = \frac{1}{\omega}\left[\Tr(\left.D_{\lat}^{(m)\,-1}\right\vert_{\cond{h^r}{h^i}})\right. \left.+ \Tr(\left.D_{\lat}^{(m)\,-1}\right\vert_{\cond{g}{h^i}}) -\frac{a}{2}\right].
\label{eq:chiralcond3}
\end{equation}

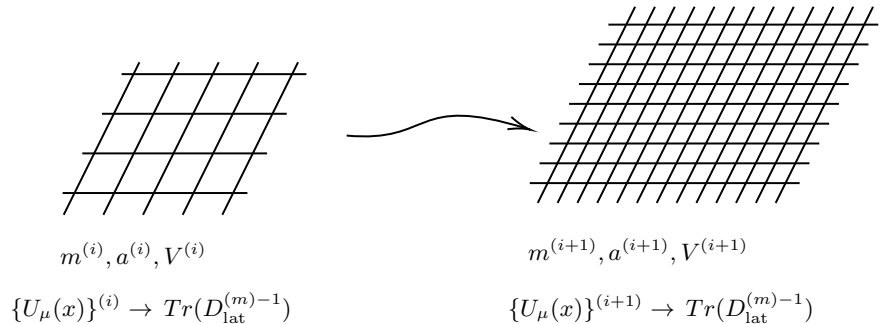
\begin{figure}[htb]
\begin{center}
\tikzset{every picture/.style={line width=0.75pt}} 
\begin{tikzpicture}[x=0.75pt,y=0.75pt,yscale=-1,xscale=1]

\draw  [draw opacity=0] (169.5,216) -- (262.69,216) -- (224.5,293) -- (131.31,293) -- cycle ; \draw   (175.5,216) -- (137.31,293)(195.5,216) -- (157.31,293)(215.5,216) -- (177.31,293)(235.5,216) -- (197.31,293)(255.5,216) -- (217.31,293) ; \draw   (166.52,222) -- (259.71,222)(156.6,242) -- (249.79,242)(146.69,262) -- (239.88,262)(136.77,282) -- (229.96,282) ; \draw    ;

\draw  [draw opacity=0] (416.5,188) -- (552.6,188) -- (503.5,287) -- (367.4,287) -- cycle ; \draw   (425.5,188) -- (376.4,287)(435.5,188) -- (386.4,287)(445.5,188) -- (396.4,287)(455.5,188) -- (406.4,287)(465.5,188) -- (416.4,287)(475.5,188) -- (426.4,287)(485.5,188) -- (436.4,287)(495.5,188) -- (446.4,287)(505.5,188) -- (456.4,287)(515.5,188) -- (466.4,287)(525.5,188) -- (476.4,287)(535.5,188) -- (486.4,287)(545.5,188) -- (496.4,287) ; \draw   (412.04,197) -- (548.14,197)(407.08,207) -- (543.18,207)(402.12,217) -- (538.22,217)(397.16,227) -- (533.26,227)(392.2,237) -- (528.3,237)(387.24,247) -- (523.34,247)(382.28,257) -- (518.38,257)(377.32,267) -- (513.42,267)(372.36,277) -- (508.46,277) ; \draw    ;

\draw    (280,253) .. controls (324.06,256.96) and (312.73,231.52) .. (370.72,249.44) ;
\draw [shift={(372.5,250)}, rotate = 197.57] [color={rgb, 255:red, 0; green, 0; blue, 0 }  ][line width=0.75]    (10.93,-3.29) .. controls (6.95,-1.4) and (3.31,-0.3) .. (0,0) .. controls (3.31,0.3) and (6.95,1.4) .. (10.93,3.29)   ;
	
	\draw (134,305.4) node [anchor=north west][inner sep=0.75pt]    {$m^{( i)} ,a^{( i)} ,V^{( i)}$};
	\draw (370,303.4) node [anchor=north west][inner sep=0.75pt]    {$m^{( i+1)} ,a^{( i+1)} ,V^{( i+1)}$};
	\draw (109,330.4) node [anchor=north west][inner sep=0.75pt]    {$\{U_{\mu}(x)\}^{(i)}\rightarrow\,Tr( D^{(m)-1}_{\lat})$};
	\draw (360,330.4) node [anchor=north west][inner sep=0.75pt]    {$\{U_{\mu}(x)\}^{(i+1)}\rightarrow\,Tr( D^{(m)-1}_{\lat})$};
	
	\end{tikzpicture}
\end{center}
\caption{Evaluating $\Tr(D^{(m)-1}_{\lat})$ for different gauge configurations generated changing parameters $m$, $a$ and $V$.}
\label{fig:1}
\end{figure}

\noindent
Now we can take $\lc$ in both sides in (\ref{eq:chiralcond3}). 
However, we need to specify the meaning of taking $\lc$ in this case. 
Indeed, it is obvious that considering such limits after the gauge average of (\ref{eq:chiralcond1})  is well defined, since 
$\langle\Sigma(m,a,V)\rangle_G$ is just a function of $m$, $a$ and $V$. 
However taking the limits just for a given gauge configuration, is less clear because we need to specify how $\Sigma(m,a,V)$ depends by $m$, $a$ and $V$ and eventually what is the meaning of taking the gauge average afterwards.
Nevertheless, the procedure adopted here for considering such limits does not differ much from how conceptually we do for quantities after we have gauge averaged. 
At first, we need to build the dependence of our quantity with respect $m$, $a$ and $V$, once we obtain it, we consider its limits, $\lc$. 
For example, we can take $\Tr(D^{(m)\,-1}_{\lat})$, which from (\ref{eq:chiralcond2}) is proportional to $\Sigma(m,a,V)$, calculated in a gauge configuration generated by a probability distribution, which is  $\propto\exp(-S)$, depending by $m$, $a$ and $V$; 
afterwards, we slight vary $m$, $a$ and $V$ and we get another probability distribution from which we generate our new gauge configuration and then we calculate $\Tr(D^{(m)\,-1}_{\lat})$ in this new gauge set. 
At the end, continuing with this procedure, varying step by step $m$, $a$ and $V$, we obtain how $\Tr(D^{(m)\,-1}_{\lat})$ depends by $m$, $a$ and $V$ and therefore we can take the limits $\lim_{\chi}$. 
In Fig. \ref{fig:1}, we summarize this procedure, where we go from a configuration generated using parameters $m^{(i)}$, $a^{(i)}$ and $V^{(i)}$ to the one generated with parameters $m^{(i+1)}$, $a^{(i+1)}$ and $V^{(i+1)}$, and we evaluate the corresponding value of $\Tr(D^{(m)\,-1}_{\lat})$ on these configurations in order to explore its dependence by $m$, $a$ and $V$. 
Moreover, the two grids in Fig. \ref{fig:1}, represent pictorially the lattice points where is defined each gauge link and they increase under the limits $a\to0$ and $V\to \infty$. 
Hence also the dimensionality of $D^{(m)\,-1}_{\lat}$ increases in such process. 
In our case the thing is simple, since the expression of its trace, given in (\ref{eq:chiralcond2}), is encoded in the expression of $\rho_a(m,V,\eta)$, from which we do not care about its structure, but only that the limit of this function exists and it does not diverge upon our limits.

Now, keeping in mind this procedure of taking $\lc$, 
we use eqs. (\ref{eq:glim})-(\ref{eq:hrlim}) and from eq. (\ref{eq:chiralcond3}), we get therefore that $\lc \Sigma (m,a,V) = (\pi/2)\rho_0(0,0)$. 
The important thing, to notice here, is that, after thermodynamic, continuum and massless limit, the only relevant contribution to the chiral condensate and, consequently, to the chiral symmetry breaking is given by the terms of the quark propagator proportional to $g(m,\eta)$. Instead the $h^i(m,\eta)$-terms (see eq. (\ref{eq:qprop2})) are traceless and $h^r(m,\eta)$ terms are zero in the continuum limit. 

Now we go back to the eq. (\ref{eq:glim}), and we assume that there is a gap in the Dirac spectrum.
Namely $\rho_a(m,V,\eta)$ is given by 

\begin{equation}
\rho_a(m,V,\eta)= \left\lbrace 
\begin{matrix}
\neq 0 & \mbox{for}\; \eta > \Lambda\\
= 0 & \mbox{for}\; \eta \leq \Lambda\\
\end{matrix}\right.
\label{eq:gap}
\end{equation}

\noindent
where $\Lambda = \Lambda(m,V,a)\geq 0 $ is the size of our gap. 
Supposing that such gap persists in the limit $\lc \Lambda >0$, then the Banks-Casher relation implies that $\rho_0(0,0) = 0$.
This has a consequence in the structure of $ D_{\lat}^{(m)\,-1}$ in the chiral limit. 
The procedure intended here, is evaluating $\lc D_{\lat}^{(m)\,-1}$, as limit of each element of the matrix $D_{\lat}^{(m)\,-1}$. 
Now, a given element depends by the gauge configuration $\{U_{\mu}(x)\}$, 
hence we can consider the above procedure which we applied for $\Tr(D_{\lat}^{(m)\,-1})$, and explore the dependence of it, just generating gauge configurations changing at each time the values of $m$, $a$ and $V$, 
in order to discover the dependence of $D_{\lat}^{(m)\,-1}$ from these parameters. 
However, without going in details on how to take this limit, we can just say which part of the quark propagator $D_{\lat}^{(m)\,-1}$, that has been decomposed in eqs. (\ref{eq:qprop2}) and (\ref{eq:qprop3}) can be negligible after $\lc$ is applied.  
For doing so, we use the result of Appendix \ref{app:trace} in eq. (\ref{eq:qpropdis2}) together with the decomposition in (\ref{eq:qprop2}), 
and we get

\begin{equation}
\lc D_{\lat}^{(m)\,-1} = \lc D_{\lat}^{(m)\,-1}\vert_{\cond{g}{h^r}} + 
\lc D_{\lat}^{(m)\,-1}\vert_{\cond{h^r}{h^i}}.
\label{eq:propgap1more}
\end{equation}

Now, if we assume a gap in the Dirac spectrum, then $\rho_0(0,0) = 0$, 
hence again from eq. (\ref{eq:qpropdis2}), we have that $\lc D_{\lat}^{(m)\,-1}\vert_{\cond{h^r}{h^i}} \underset{\gap}{=} 0 $, 
where we denoted with $\underset{\gap}{=}$, such gap assumption in the eigenvalue distribution given in eq. (\ref{eq:gap}). Therefore we have

\begin{equation}
\lc D_{\lat}^{(m)\,-1} \underset{\gap}{=} \lc \left. D_{\lat}^{(m)\,-1}\right\vert_{\cond{g}{h^r}},
\label{eq:propgap1}
\end{equation} 

Hence only the $h^i$-term becomes relevant, which allows us to define the quark propagator:

\begin{equation}
D_{\gap}^{-1} \underset{\gap}{=} \lc D_{\lat}^{(m)\,-1}.
\label{eq:propgap2}
\end{equation}

\noindent
We can also give, explicitly, the matrix structure of $D_{\gap}^{-1}$. 
Take, in fact, the Dirac structure of the Dirac eigenvectors $v_n^{(\pm)}$:

\begin{equation}
v_n^{(\pm)} = \left(\begin{matrix}
\pm\, L_{n}\\
R_{n}
\end{matrix}\right),
\label{eq:eigvect}
\end{equation}

\noindent
where we used $\gamma_5$ defined in (\ref{eq:gamma1}).
Then, from the expression of $D_{\gap}^{-1}\vert_{\cond{g}{h^r}}$, $D_{\gap}^{-1}$ can be written as

\begin{equation}
D_{\gap}^{-1} = \lc\sum_{\eta_n>\Lambda} \I h^i(m,\eta_n)\left(\begin{matrix}
0 & L_n R_n^{\dagger}\\
R_n L_n^{\dagger} & 0 
\end{matrix}\right),
\label{eq:qprop4}
\end{equation}

\noindent
where in the sum we explicitly imposed that, there is a gap in the Dirac spectrum of size $\Lambda$. 
Moreover, directly from the matrix structure in eq. (\ref{eq:qprop4}), we have that $D_{\gap}^{-1}$ satisfies the relations

\begin{equation}
\{\gamma_5, D_{\gap}^{-1}\} = 0,\quad \Tr(D_{\gap}^{-1}) = 0,
\label{eq:gamma5anticomm}
\end{equation}

\noindent
which are pretty trivial to prove. 
Indeed, from eqs. (\ref{eq:propgap1}) and (\ref{eq:propgap2}), 
we have $D_{\gap}^{-1} = \lc D_{\lat}^{(m)\,-1}\vert_{\cond{g}{h^r}}$. 
Now, $\gamma_{5}D_{\gap}^{-1} \gamma_{5} = \lc (\gamma_{5} D_{\lat}^{(m)\,-1}\vert_{\cond{g}{h^r}}\gamma_{5})$, 
however $\gamma_{5}D_{\lat}^{(m)\,-1}\vert_{\cond{g}{h^r}} \gamma_{5}=-D_{\lat}^{(m)\,-1}\vert_{\cond{g}{h^r}}$, that can be easy checked from its structure in (\ref{eq:qprop3}). 
Hence $\gamma_{5}D_{\gap}^{-1} \gamma_{5} = -\lim_{\chi}D_{\lat}^{(m)\,-1}\vert_{\cond{g}{h^r}} = -D_{\gap}^{-1} $ and this proves the first equality in (\ref{eq:gamma5anticomm}). 
The second one uses that $\Tr(D_{\gap}^{-1}) = 
\Tr(\lc D_{\lat}^{(m)\,-1}\vert_{\cond{g}{h^r}}) = \lc \Tr(D_{\lat}^{(m)\,-1}\vert_{\cond{g}{h^r}}) = 0$, because of eq. (\ref{eq:qproptrace1}).

\section{Axial and chiral symmetry}\label{sec:axial}

Once we have made the hypothesis of a gap in the Dirac spectrum (see eq. (\ref{eq:gap})), we want to show how the propagator $D_{\gap}^{-1}$ transforms under $U(1)_A$ and $SU(n_F)_L\times SU(n_F)_R$ groups. 
As we said in section \ref{sec:prelim}, we consider the simple case in which no zero modes are present in the theory and no 't Hooft term is present in the action. 

At first we notice that, the quark propagator $D_{\gap}^{-1}$ can be also written as $
D_{\gap}^{-1} \underset{\gap}{=} \lc\, \langle \psi_i \bar{\psi}_i\rangle_F$, 
where we basically used eq. (\ref{eq:propgap2}) and the fact that $D_{\lat}^{(m)\,-1} = \langle \psi_i \bar{\psi}_i\rangle_F$, with $i$ a fixed flavor index. 
Now, looking how the fields $\psi_i $ and $\bar{\psi}_i$ transform under $U(1)_A$ and $SU(n_F)_L\times SU(n_F)_R$ groups, we can see how $D_{\gap}^{-1}$, transforms.

Starting from $U(1)_A$ transformations, they are implemented as

\begin{equation}
U(1)_A:\;\psi_i\To \exp{(\I\alpha\gamma_5)}\psi_i,\quad\bar{\psi}_i\To \bar{\psi}_i\exp{(\I\alpha\gamma_5)},
\label{eq:u1atrans1}
\end{equation}

\noindent
with $\gamma_5$ given in eq. (\ref{eq:gamma1}).
Typically on the lattice the $\gamma_5$ matrix is replaced by $\hat{\gamma}_5 = \gamma_5(\mathds{1} - (a/2)D_{\lat}^{-1})$, which includes a correction of finite lattice spacing. 
However for simplicity, we consider the standard $\gamma_5$, since whatever correction we apply, should vanish anyhow in the continuum limit. 

Inserting the transformations (\ref{eq:u1atrans1}) in $D_{\gap}^{-1}$, we get

\begin{equation}
\begin{split}
U(1)_A:\;D_{\gap}^{-1}\To D_{\gap}^{-1\,U(1)_A} = \cos^2(\alpha)\, D_{\gap}^{-1} + \I \sin(\alpha)\,\cos(\alpha)\,(\gamma_5D_{\gap}^{-1}+D_{\gap}^{-1} \gamma_5) - \sin^2(\alpha)\, \gamma_5 D_{\gap}^{-1}\gamma_5.
\end{split}
\label{eq:u1atrans2}
\end{equation}

\noindent
Using now the anticommutation relation in eq. (\ref{eq:gamma5anticomm}) for $D_{\gap}^{-1}$, we get that $D_{\gap}^{-1}$ is invariant under $U(1)_A$ transformations, namely
$	D_{\gap}^{-1\,U(1)_A} = D_{\gap}^{-1}$.
This result is quite expected, if you consider that we are not taking in consideration the eventual transformation of the measure $\mathcal{D}\left[ \psi \bar{\psi}\right]$ in $\langle \psi_i \bar{\psi}_i\rangle_F$ under $U(1)_A$. 

Now we look the transformation of $(D_{\gap}^{-1})_{ij} = \lc\, \langle \psi_i \bar{\psi}_j \rangle$ (where $i,j$ are generic flavor indices) under the chiral group $SU(n_F)_L\times SU(n_F)_R$. 

At first we need to remind again our assumptions. 
As we already said in section \ref{sec:prelim}, we suppose that the fermionic action can be split into the sum of the action of different quark flavors, i.e. $S_F = \sum_{i=1}^{n_F} S_i$, then the mean value $\langle \psi_i \bar{\psi}_j \rangle$ turns to be zero for $i\neq j$, because $\langle \psi_i \bar{\psi}_j \rangle = \langle \psi_i\rangle\langle \bar{\psi}_j \rangle = 0$, since $\langle \bar{\psi}_j \rangle$ and $\langle \psi_i \rangle$ are null for all $i,j$. 
In this case we can write that 
$(D_{\gap}^{-1})_{ij} = \delta_{ij}D_{\gap}^{-1} $, with $D_{\gap}^{-1}= \lc\, \langle \psi_i \bar{\psi}_i \rangle$ for some given flavor $i$.

We are now ready for our calculations.
Defining the quark fields $\psi_{L/R} = P_{L/R} \psi$ and $\bar{\psi}_{L/R} = \bar{\psi}_{L/R}P_{R/L}$, with $P_{L} = (\mathds{1}-\gamma_5)/2$ and $P_{R} = (\mathds{1}+\gamma_5)/2$, then the chiral transformations are implemented as

\begin{equation}
\begin{split}
&\psi_{L} \To \exp(\I\alpha_{L}^a T^a)\psi_{L},\quad\bar{\psi}_{L} \To \bar{\psi}_{L}\exp(-\I\alpha_{L}^a T^a),\quad\psi_{R} \To \exp(\I\alpha_{R}^a T^a)\psi_{R},\quad\bar{\psi}_{R} \To \bar{\psi}_{
	R}\exp(-\I\alpha_{R}^a T^a),\\
\end{split}
\label{eq:LRspinor1}
\end{equation}

\noindent
where $T^a$ are the generators of the group $SU(n_F)$ . 
Using that $\psi = \psi_L + \psi_R$ and $\bar{\psi} = \bar{\psi}_L + \bar{\psi}_R$, we have that $(D_{\gap}^{-1})_{ij}$ can be decomposed as (see Appendix \ref{app:suLR})

\begin{equation}
\begin{split}
(D_{\gap}^{-1})_{ij} = (D_{LL}^{-1})_{ij} + (D_{RR}^{-1})_{ij},
\end{split}
\label{eq:qpropRL}
\end{equation}

\noindent
where $(D_{LL}^{-1})_{ij} = \lc\, \langle \psi_{L,i} \bar{\psi}_{L,j} \rangle$, $(D_{RR}^{-1})_{ij} = \lc\, \langle \psi_{R,i} \bar{\psi}_{R,j} \rangle$.

The eq. (\ref{eq:qpropRL}) tells us that in presence of a gap only the $LL$ and $RR$ components survive in the quark propagator. 
Other possible terms like $\lc\, \langle \psi_{L,i} \bar{\psi}_{R,j}\rangle$ and $\lc\, \langle \psi_{R,i} \bar{\psi}_{L,j}\rangle$ are zero, because of eq. (\ref{eq:gamma5anticomm}), as we show in Appendix \ref{app:suLR}. 

It is now straightforward proving that $(D_{LL}^{-1})_{ij}$ and $(D_{RR}^{-1})_{ij}$ are invariant under the transformations in (\ref{eq:LRspinor1}). 
Indeed, from the previous discussion, we can write that, $(D_{XX}^{-1})_{ij} = \delta_{ij}D_{XX}^{-1}$ for $X = L$ or $R$, with $D_{XX}^{-1} = \lc\, \langle \psi_{X,i} \bar{\psi}_{X,i}\rangle $ for a given index flavor $i$. 
Therefore under $SU(n_F)_L\times SU(n_F)_R$ transformations we have

\begin{equation}
\begin{split}
&(D_{XX}^{-1})_{ij}\To (D_{XX}^{-1})_{ij}^{SU(n_F)_X}=(\exp(\I\alpha_{X}^a T^a))_{im} (D_{XX}^{-1})_{mn}(\exp(-\I\alpha_{X}^a T^a))_{nj}\\
&= (\exp(\I\alpha_{X}^a T^a))_{im} \delta_{mn}(D_{XX}^{-1})(\exp(-\I\alpha_{X}^a T^a))_{nj}=\delta_{ij}(D_{LL}^{-1}) = (D_{XX}^{-1})_{ij}.
\end{split}
\label{eq:qpropXX}
\end{equation}

\noindent
Hence from eq. (\ref{eq:qpropRL}) we get

\begin{equation}
\begin{split}
&(D_{\gap}^{-1})_{ij}\To (D_{\gap}^{-1})_{ij}^{SU(n_F)_L \times SU(n_F)_R} = (D_{RR}^{-1})_{ij}^{SU(n_F)_R} +
(D_{LL}^{-1})_{ij}^{SU(n_F)_L} = (D_{\gap}^{-1})_{ij},
\end{split}
\label{eq:qpropRLt}
\end{equation}

\noindent
where we used eq. (\ref{eq:qpropXX}).
Therefore $D_{\gap}^{-1}$ is invariant under $SU(n_F)_L\times SU(n_F)_R$ group, when a gap open in the Dirac spectrum. 

These results are expected, because we assumed an eigenvalue distribution as in eq. (\ref{eq:gap}), with some gap $\Lambda$. 
The reason is that, we have basically looked the Banks-Casher relation from the other way around. If $\rho_0(0,0) = 0$ and no zero modes are present in the theory, then $U(1)_A$ and $SU(n_F)_L\times SU(n_F)_R$ are restored in the massless limit. 

\section{Symmetries in correlators}\label{sec:corr}

We look now how the invariance of $D_{\gap}^{-1}$ under $U(1)_A$ and $SU(n_F)_L \times SU(n_F)_R$ implies also the invariance of the hadron correlators. 

A general observable, which is a function of a set of variables $\{\psi\}$ and $\{\bar{\psi}\}$, in different indices, can be written as

\begin{equation}
\mathcal{C}(\psi,\bar{\psi}) = 
\sum_{k}\sum_{I_1,...,I_{2k}}\acorrkmin \prod_{l=1}^{k}\psi(x_l)_{I_l}\bar{\psi}(x_{k+l})_{I_{k+l}},
\label{eq:corr1}
\end{equation}

\noindent
where the $\Gamma$s represent different coefficients, $I_{l} = \{i_l,\alpha_{l},a_{l}\}$, for $l=1,...,2k$, represents a multi-index, which contains flavor, Dirac and color indices respectively. 
In the sense that, for a given $l$, then $i_l$ and $i_{k+l}$ go from $1$ to $n_F$ (number of flavors), $\alpha_l$ and $\alpha_{k+l}$ go from $1$ to $4$ (number of Dirac indices), and $a_l$ and $a_{k+l}$ go from $1$ to $N_c$ (number of colors).
In eq. (\ref{eq:corr1}), we have put the same number of $\psi$ and $\bar{\psi}$ variables, because otherwise $\langle\mathcal{C}(\psi,\bar{\psi})\rangle_F$ would be zero, for the Wick theorem and the fact that $\langle\psi\rangle_F =\langle \bar{\psi}\rangle_F = 0$.
Taking $\lc\langle\mathcal{C}(\psi,\bar{\psi})\rangle_F$ (with our specifications on such limit given in section \ref{sec:gap}), we can write it in terms of the quark propagator. 
As it is shown in Appendix \ref{app:corr}, under the assumption of a gap in the Dirac spectrum (\ref{eq:gap}), $\mathcal{C}(\psi,\bar{\psi})$ can be written as

\begin{equation}
\begin{split}
&\lc\, \langle\mathcal{C}(\psi,\bar{\psi})\rangle_F \underset{\gap}{=} \sum_{k}\sumcorrmin\sum_p s(p)\, \acorrkbarmin\prod_{l=1}^k\delta_{p(i_l)i_{k+l}} D^{-1}_{\gap}(x_{p(l)},x_{k+l})_{p(\alpha_{l})p(a_l),\alpha_{k+l} a_{k+l}},
\end{split}
\label{eq:corr2}
\end{equation}	

\noindent
where we have put $D^{-1}_{\gap}$, given in eq. (\ref{eq:qprop4}).
In eq. (\ref{eq:corr2}), $\bar{\Gamma} = \lc \Gamma$, instead $p(l)$ (as well as $p(\alpha_{l})$ and $p(a)$) is the label obtained after the application of $p$ transpositions of $l$ (respectively $\alpha$ and $a$), and we are summing over all possible transpositions $p$; $s(p)$ is the sign which each permutation gives, coming from the exchange of $I_l$ and $p(I_l)$.
The Kronecker delta $\delta_{p(i_l)j_l}$ indicates that for different quark flavors the contribution is zero, since we are assuming that no interaction terms among quark flavors is present in the fermionic action.

The eq. (\ref{eq:corr2}) tells us that a generic correlator can be written as a linear combination of the quark propagator $D^{-1}_{\gap}$.
Therefore under transformations $\psi \To \psi^{\mathcal{G}}$ and $\bar{\psi}\To\bar{\psi}^{\mathcal{G}}$, with $\mathcal{G} = U(1)_A$ or $SU(n_f)_L\times SU(n_f)_R$, we have that

\begin{equation}
\begin{split}
\lc\, \langle\mathcal{C}(\psi^{\mathcal{G}},\bar{\psi}^{\mathcal{G}})\rangle_F &\underset{\gap}{=}\sum_{k}\sumcorrmin\sum_p s(p)\, \acorrkbarmin\prod_{l=1}^k\delta_{p(i_l)i_{k+l}} D^{-1}_{\gap}(x_{p(l)},x_{k+l})_{p(\alpha_{l})p(a_l),\alpha_{k+l} a_{k+l}}^{\mathcal{G}}\\
&=\sum_{k}\sumcorrmin\sum_p s(p)\, \acorrkbarmin\prod_{l=1}^k\delta_{p(i_l)i_{k+l}} D^{-1}_{\gap}(x_{p(l)},x_{k+l})_{p(\alpha_{l})p(a_l),\alpha_{k+l} a_{k+l}}\\
&= \lc\, \langle\mathcal{C}(\psi,\bar{\psi})\rangle_F,
\end{split}
\label{eq:corr3}
\end{equation}	

\noindent
where in second line, we used the results of section \ref{sec:axial}, which tell us that $D_{\gap}^{-1\,\mathcal{G}} = D_{\gap}^{-1}$.
Now if the distribution of the Dirac eigenvalues has a gap for all gauge configurations, then the symmetry in eq. (\ref{eq:corr3}) is held 
for each gauge configuration, especially after an arithmetic mean.
Namely, $\lc\frac{1}{N}\sum_{i=1}^N \langle\mathcal{C}(\psi,\bar{\psi})\rangle_F^{(i)} = \frac{1}{N}\sum_{i=1}^N \lc\langle\mathcal{C}(\psi,\bar{\psi})\rangle_F^{(i)} = 
\frac{1}{N}\sum_{i=1}^N\lc \langle\mathcal{C}(\psi^{\mathcal{G}},\bar{\psi}^{\mathcal{G}})\rangle_F^{(i)}
= \lc
\frac{1}{N}\sum_{i=1}^N \langle\mathcal{C}(\psi^{\mathcal{G}},\bar{\psi}^{\mathcal{G}})\rangle_F^{(i)}
$, 
because every term inside the sum for $i=1,...,N$ satisfies (\ref{eq:corr3}). 
From the lattice simulation perspective, we use the arithmetic mean as an estimator for the gauge average, i.e. 
we generally approximate $\langle\langle\mathcal{C}(\psi,\bar{\psi})\rangle_F\rangle_G$ $\approx\frac{1}{N}\sum_{i=1}^N \langle\mathcal{C}(\psi,\bar{\psi})\rangle_F^{(i)}$. 
Therefore we can ask a question whether $\lc\langle\langle\mathcal{C}(\psi,\bar{\psi})\rangle_F\rangle_G= 
\lc\langle\langle\mathcal{C}(\psi^{\mathcal{G}},\bar{\psi}^{\mathcal{G}})\rangle_F\rangle_G$, 
since we have seen that substituting $\langle\cdot\rangle_G\to\frac{1}{N}\sum_{i=1}^N \cdot$, such equation is held. 
Unfortunately, we cannot answer with a sure statement on this, 
because it depends how the arithmetic mean gives a good estimation of the gauge average 
and a more rigorous approach is needed in this case. 
However for practical purposes, in lattice simulations the gauge averages are always taken as arithmetic mean, 
therefore it seems to be reasonable finding chiral and axial symmetry effectively restored in hadron correlators upon a gap in the Dirac spectrum.

\section{Chiralspin symmetry and eigenvectors}\label{sec:cs}

Chiralspin symmetry, seems to emerge in the hadron spectrum when the lowest eigenmodes of the Dirac operator are suppressed. This can happen surgically, by removing them from the quark propagator see \cite{Denissenya:2014poa,Denissenya:2015mqa,Denissenya:2015woa}, 
or at high temperature ($T>1.2\, T_c$) as shown in \cite{Rohrhofer:2017grg,Rohrhofer:2019qal,Rohrhofer:2019qwq}, where a gap in the Dirac spectrum as in eq. (\ref{eq:gap}) appears naturally, see \cite{Bazavov:2012qja,Cossu:2013uua,Tomiya:2016jwr} for that. 
Therefore, it seems that the necessary condition for having the $SU(2)_{CS}$ symmetry in the hadron spectrum is that, there is a gap in the eigenvalue distribution, whenever if this happens manually or naturally at high temperature. 

Such symmetry appears basically as a degeneration of hadron masses and, more in general, in hadron correlators summed over the space indices and then correctly normalized with the space volume. 
However, doing this sum analytically, in the most general case, can be difficult. 
Therefore we adopt a different procedure. 
Suppose to have two generic correlators of some hadrons connected via $SU(2)_{CS}$ group, namely $C_1(x,y) = \langle \mathcal{O}_1(x)\bar{\mathcal{O}}_1(y)\rangle$ and $C_2(x,y)= \langle \mathcal{O}_2(x)\bar{\mathcal{O}}_2(y)\rangle$, summed over $\bm{x}$ and $\bm{y}$, i.e. $C_i(t) = \sum_{\bm{x},\bm{y}}C_i(x,y)$, for $i=1,2$. 
In this case for large $t = \vert x_4 - y_4 \vert$, the dominant part is given by the exponential of the mass, i.e. $C_i(t)\sim \exp(-m_i\,t)$, with $m_i$ with $i=1,2$, the masses of the two given hadrons. 
Therefore in such limit, the degeneration of the masses $m_i$ corresponds to a degeneration of $C_i(t)$ and vice versa. 
We can take now the following temporal correlators $C_i (x_4,y_4)= \langle \mathcal{O}_i(x_4)\bar{\mathcal{O}}_i(y_4)\rangle$ for $i=1,2$, where we have set $\bm{x}$ and $\bm{y}$ to zero. 
Then, also in this case for large $t = \vert x_4 - y_4 \vert$, we can write that $C_i (x_4,y_4)\sim\exp(-m_i\,t)$. 
Hence even in this case, a degeneration of the masses $m_i$ correspond to a degeneration of $C_i(x_4,y_4)$ and vice versa, when $t$ is large enough. 
Now, the hadron correlators $C_i(x_4,y_4)$ are basically equal to the ones given in eq. (\ref{eq:corr1}), but with a proper choice of the coefficients $\Gamma$ and averaged over the fermionic and the gluon action, i.e. $\langle\cdot\rangle = \langle\langle\cdot\rangle_F\rangle_G$. 
Such correlators can be always expressed in terms of the quark propagator as we have seen in section \ref{sec:corr}. 
In particular $C_i(x_4,y_4)$, upon the limits $ \lim_{m\To 0}\lim_{a\To 0}\lim_{V\To\infty}$ and the gap hypothesis (\ref{eq:gap}), can be expressed in terms of the quark propagator $D_{\gap}^{-1}(x_4,y_4)$. 
Therefore a symmetry of the quark propagator $D_{\gap}^{-1}(x_4,y_4)$ induces a symmetry of $C_i(x_4,y_4)$ and consequently a degeneration of the hadron masses. 
Hence, here we want to conjecture that the symmetry $SU(2)_{CS}$ arises from an invariance of $D_{\gap}^{-1}$, similar on what we have observed in the case of axial and chiral symmetry in section \ref{sec:axial}. 
In particular we want to impose the $SU(2)_{CS}$ symmetry in $D_{\gap}^{-1}(x_4,y_4)$, which is the quark propagator in the time direction.

In order to impose such symmetry, we observe that

\begin{equation}
D_{\gap}^{-1}(x_4,y_4)\underset{\gap}{=} \lc\langle\psi(x_4)_i\bar{\psi}(y_4)_i\rangle_F,
\label{eq:qprop5}
\end{equation}

\noindent
for a given flavor $i$. 
Hence for the calculation of $D_{\gap}^{-1}(x_4,y_4)$ we just need to consider the quark fields in the points $x = (x_4,\bm{0})$ and $y = (y_4,\bm{0})$.

Before to proceed in our calculations, we remind what the $SU(2)_{CS}$ group is. 
The $SU(2)_{CS}$ group is defined by the set of transformations of the quark fields generated by $\bm{\Sigma}=\{\gamma_4,\I\gamma_5\gamma_4,-\gamma_5\}$, which forms an $\mathrm{su}(2)$ algebra. 
Taking two generic quark fields $\psi(x)_i$ and $\bar{\psi}(x)_i$ for a given flavor index $i$, in the point $x = (x_4,\bm{0})$, the $SU(2)_{CS}$ transformations are given by 

\begin{equation}
\begin{split}
&\psi(x_4)_i\To\exp(\I\alpha_a \Sigma_a)\,\psi(x_4)_i,\quad\bar{\psi}(x_4)_i\To\bar{\psi}(x_4)_i\,\gamma_4\exp(-\I\alpha_a \Sigma_a)\gamma_4
\end{split}
\label{eq:su2cs1}
\end{equation}

\noindent
where we used the definition of chiralspin transformations as in \cite{Glozman:2015qva}. 
We can now plug the transformations (\ref{eq:su2cs1}) in eq. (\ref{eq:qprop5}) and look how $D_{\gap}^{-1}(x_4,y_4)$ transforms. 
This is done in Appendix \ref{app:su2cs}. 
We found that $D_{\gap}^{-1}(x_4,y_4)$ is not invariant under $SU(2)_{CS}$, but only under its subgroup $U(1)_A \subset SU(2)_{CS}$. 
Hence in Appendix \ref{app:su2cs}, we also give the condition that $D_{\gap}^{-1}(x_4,y_4)$ needs to satisfy in order to be chiralspin symmetric.
Defining as $D^{-1}_{\gap,\CS}$ the chiralspin symmetric quark propagator, then 
the condition of chiralspin invariance of the quark propagator is

\begin{equation}
\gamma_4 D^{-1}_{\gap,\CS}(x_4,y_4)\gamma_4 = D^{-1}_{\gap,\CS}(x_4,y_4), 
\label{eq:cond1}
\end{equation}

\noindent
which is shown in Appendix \ref{app:su2cs}.
Using now the expression (\ref{eq:eigvect}) for the eigenvectors and  (\ref{eq:qprop4}), we obtain that the sufficient condition for satisfying
eq. (\ref{eq:cond1}) is that

\begin{equation}
L_n(x_4) R_n^{\dagger}(y_4) \underset{\CS}{=} R_n(x_4) L_n^{\dagger}(y_4).
\label{eq:cond2}
\end{equation}

\noindent
From this equation, we can give the following \textit{ansatz} on the eigenvector structure

\begin{equation}
v^{(\pm)}_n(x_4) \underset{\CS}{=} \left(\begin{matrix}
\pm\chi_n(x_4)\\
\tau_n \chi_n(x_4)
\end{matrix}\right),\quad\mbox{with}\quad \tau_n^* = \tau_n
\label{eq:eigstr1}
\end{equation}

\noindent
where $\chi_n(x)$ is some generic $2$-component field and $\tau_n$ is a real operator. 
Finally the quark propagator looks like 

\begin{equation}
D^{-1}_{\gap,\CS}(x_4,y_4) = \lc\sum_{\eta_n>\Lambda} \I h^i(m,\eta_n)\,\tau_n\chi_n(x_4)\chi_n^{\dagger}(y_4)\gamma_4,
\label{eq:qpropsu2cs1}
\end{equation}

\noindent
which is our $SU(2)_{CS}$-invariant quark propagator in $x = (x_4,\bm{0})$ and $y = (y_4,\bm{0})$. 

Now the reason why chiralspin symmetry emerges when there is a gap in the Dirac spectrum is difficult to understand. 
It may be given by dynamical reasons as pointed out in \cite{Catillo:2019jrl} and connected by a stringy fluid matter structure, as explained in \cite{Glozman:2018jkb}. 
Here we have just imposed the chiralspin symmetry in the temporal quark propagator in order to get the structure of the Dirac eigenvectors (\ref{eq:eigstr1}) at $x = (x_4,\bm{0})$. 
If such quark propagator is implemented in the calculation of the temporal correlators $C(x_4,y_4)$, it will induce the $SU(2)_{CS}$ symmetry in the hadron spectrum.
However, we conclude saying that the gap in the Dirac spectrum is just a necessary condition for having such symmetry, but still not sufficient and other input are needed. 
This is also remarked by the fact that at $T> 3\,T_c$, the chiralspin symmetry seems to disappear, as pointed out in recent works \cite{Rohrhofer:2019qwq,Rohrhofer:2019qal}. 
In such regime the gap in the Dirac spectrum persists, but chiralspin symmetry does not.

\section{Conclusions}\label{sec:conclusion}

We can summarize, now, our main results. 
We have started from a lattice formulation of the quark propagator and we have seen that when there is a gap in the distribution of the Dirac eigenvalues and we are not in presence of anomaly, then, under the limits $\lim_{m\To0}\lim_{a\To0}\lim_{V\To\infty}$, we have that the quark propagator simplifies as in eq. (\ref{eq:qprop4}) and it becomes invariant under $SU(n_F)_L \times SU(n_F)_R$ and $U(1)_A$ transformations. 
This induces an invariance of whatever observable which is a function of the quark fields, especially the hadron correlators, bringing to a degeneration of the hadron masses connected through $SU(n_F)_L \times SU(n_F)_R$ and $U(1)_A$.

Moreover, upon the lattice evidence \cite{Denissenya:2014poa,Denissenya:2015mqa,Denissenya:2015woa,Rohrhofer:2017grg,Rohrhofer:2019qal,Rohrhofer:2019qwq} of a new symmetry group (namely $SU(2)_{CS}$) when the low-lying Dirac eigenmodes are suppressed (by hand or going at high temperature, $T >1.2\, T_c$), we have studied if the quark propagator, in eq. (\ref{eq:qprop4}), is also invariant under $SU(2)_{CS}$. 
We found that the only gap is not sufficient for the evidence of such symmetry. However we have imposed which condition the quark propagator, in the time direction, needs to satisfy in order to be $SU(2)_{CS}$-invariant. 
This is given in eq. (\ref{eq:cond1}) and we give also the structure of such quark propagator in eq. (\ref{eq:qpropsu2cs1}), arguing that such kind of quark propagator can lead to a degeneration of the hadron masses connected via $SU(2)_{CS}$.

\begin{acknowledgements}
I am thankful to L. Glozman and C. B. Lang for introducing me on this topic. 
I also thank Marina Marinkovi\'{c} for the support.
This work is supported by the Institute for Theoretical Physics, ETH Zurich.
\end{acknowledgements}

\appendix

\section{Conventions}\label{app:conventions}
Here we present the main conventions and notations used in this paper.

The gamma matrices in euclidean space-time are taken in the following representation:

\begin{equation}
\gamma_{\mu} = \left(\begin{matrix}
0 & \bar{\sigma}_{\mu}\\
\sigma_{\mu} & 0 
\end{matrix}\right),\qquad
\gamma_5 = \left(\begin{matrix}
-\mathds{1} & 0 \\
0 & \mathds{1}
\end{matrix}\right),
\label{eq:gamma1}
\end{equation}

\noindent
where $\sigma_{\mu} = (\mathds{1},\I\bm{\sigma})$, $\bar{\sigma}_{\mu} = (\mathds{1},-\I\bm{\sigma})$, for $\mu=1,2,3,4$, while 
$\bm{\sigma}\equiv \sigma_{1,2,3}$ are the Pauli matrices. 
The matrices in eq. (\ref{eq:gamma1}) satisfy the properties: $\{\gamma_{\mu},\gamma_{\nu}\} = 2\delta_{\mu\nu}\mathds{1}$ and $\{\gamma_{\mu},\gamma_5\} = 0$, for all $\mu,\nu$.

In this paper, vectors are indicated without indices, however you need to keep in mind that their indices structure is given by $v \equiv v(x)_{\alpha a} $, where $v$ is a generic given vector and $\alpha=1,...,4$ is the Dirac index, $a = 1,...,N_c$ is the color index, and finally $x$ is the space-time position. 
The same notation is also applied for matrices, namely given a generic matrix $A$, we have $A \equiv A(x,y)_{\alpha a,\beta b}$, with $\alpha, \beta$ Dirac indices, $a,b$ color indices and $x,y$ are two space-time points.

The scalar product is defined as $
(v,Aw) = \sum_{x,y,\alpha,\beta,a,b}
v(x)_{\alpha a}^{\dagger}A(x,y)_{ \alpha a,\beta b}\,w(y)_{\beta b}$.
In case where the matrix $A$ can be written as $A = h\,z^{\dagger}$, where $h$ and $z$ are two generic vectors, we have that $
\Tr(A) = \sum_{x,\alpha,a}h(x)_{\alpha a}z(x)_{\alpha a}^{\dagger} = (z,h)$.
This last relation is often used in section \ref{sec:gap} for the quark propagator.

\section{Traces of quark propagator}\label{app:trace}

In this appendix we derive a few relations which relate the trace of the quark propagator and its parts, with the distribution of the Dirac eigenvalues. 

At first we observe that the orthogonality relation of the Dirac eigenvectors, defined in eq. (\ref{eq:eig1}), namely $(v^{(\pm)}_n,v^{(\pm)}_m) = (1/V)\delta_{nm}$, where the scalar product is defined in section \ref{app:conventions}, implies that 
\begin{equation}
\Tr(v^{(\pm)}_n v^{(\pm)\,\dagger}_n)= (v^{(\pm)}_n,v^{(\pm)}_n) =\sum_{x,\alpha,a}\vert v^{(\pm)}_n(x)_{\alpha a}\vert^2 =1/V,
\label{eq:scalprod3}
\end{equation}

where $\vert\cdot\vert$ is simply the modulus of complex numbers
Therefore taking the trace in left-hand side of the equations in  (\ref{eq:qprop3}), we get 

\begin{equation}
\begin{split}
&\Tr(\left.D_{\lat}^{(m)\,-1}\right\vert_{\cond{h^r}{h^i}}) = \frac{2}{V} \sum_n g(m,\eta_n),\quad\Tr(\left.D_{\lat}^{(m)\,-1}\right\vert_{\cond{g}{h^i}}) = \frac{2}{V} \sum_n h^r(m,\eta_n),\quad\Tr(\left.D_{\lat}^{(m)\,-1}\right\vert_{\cond{g}{h^r}}) = 0,\\
\end{split}
\label{eq:qproptrace1}
\end{equation}

\noindent
and consequently from eq. (\ref{eq:qprop2}), 
$
\Tr(D_{\lat}^{(m)\,-1} ) = \frac{2}{V}\sum_n (g(m,\eta_n) + h^r(m,\eta_n))$.

Considering eq. (\ref{eq:scalprod3}), we have that 
each term inside the sum, has to be less than $1/V$, i.e. $\vert v_n^{(\pm)}(x)_{\alpha a} \vert^2 \leq 1/V $, for each $x$, $a$ and $\alpha$. 
Hence a simple triangle inequality gives us that $\vert v_n^{(+)}(x)_{\alpha a} v_n^{(+)}(y)_{\beta b}^{\dagger}\pm v_n^{(-)}(x)_{\alpha a} v_n^{(-)}(y)_{\beta b}^{\dagger}\vert \leq \vert v_n^{(+)}(x)_{\alpha a}\vert \vert v_n^{(+)}(y)_{\beta b}^{\dagger}\vert+\vert v_n^{(-)}(x)_{\alpha a}\vert \vert v_n^{(-)}(y)_{\beta b}^{\dagger}\vert \leq 2/V$.
This fact, together with eq. (\ref{eq:qproptrace1}), gives the following inequalities:

\begin{equation}
\begin{split}
&\left\vert \left.D_{\lat}^{(m)\,-1}(x,y)_{\alpha a, \beta b}\right\vert_{\cond{h^r}{h^i}} \right\vert\leq \Tr(\left.D_{\lat}^{(m)\,-1}\right\vert_{\cond{h^r}{h^i}}),\quad\left\vert \left.D_{\lat}^{(m)\,-1}(x,y)_{\alpha a, \beta b}\right\vert_{\cond{g}{h^i}} \right\vert\leq\Tr(\left.D_{\lat}^{(m)\,-1}\right\vert_{\cond{g}{h^i}}),
\end{split}
\label{eq:qpropdis1}
\end{equation}

\noindent
for generic indices $x,y,\alpha,\beta,a,b$. 
The inequalities in eq. (\ref{eq:qpropdis1}) are obtained from eq. (\ref{eq:qprop3}).

Now we want to use the definition of the eigenvalue distribution in eq. (\ref{eq:distrrho1}), where the eigenvalues $\eta_n = 2/a$ has been removed by hand, and see the relation with the above traces of the quark propagator. 
At first we need to consider that 

\begin{equation}
\begin{split}
& \frac{2}{V} \sum _{n:\eta_n\neq2/a} g(m,\eta_n) = 
\Tr(\left.D_{\lat}^{(m)\,-1}\right\vert_{\cond{h^r}{h^i}}) - \frac{a}{2}\frac{l}{V},\quad \frac{2}{V} \sum _{n:\eta_n\neq2/a} h^r(m,\eta_n) = 
\Tr(\left.D_{\lat}^{(m)\,-1}\right\vert_{\cond{g}{h^i}}) - \frac{a}{2}\frac{l}{V},\\
\end{split}
\label{eq:qproptrace3}
\end{equation}

\noindent
where $l$ is the multiplicity of the eigenvalues $2/a$. 
The terms $(a\,l/2V)$ become irrelevant in the thermodynamic ($V\To\infty$) and continuum ($a\To 0 $) limit.
Hence for a generic function $f(m,\eta) $ (which can be $g(m,\eta)$ or $h^r (m,\eta)$), we have, 

\begin{equation}
\begin{split}
&\frac{2}{V} \sum _{n:\eta_n\neq2/a} f(m,\eta_n) = \frac{2}{V} \int_0^{\infty}d\eta\,\sum _{n:\eta_n\neq2/a} f(m,\eta) \delta(\eta-\eta_n)= \int_0^{\infty}\rho_a(m,V,\eta)f(m,\eta),\\
\end{split}
\label{eq:rhotrace}
\end{equation}

\noindent
where we used that $\int_0^{\infty}\delta(\eta - \eta_n) = 1$ and the definition in eq. (\ref{eq:qproptrace3}). 
Therefore from eq. (\ref{eq:rhotrace}) and using eq. (\ref{eq:qproptrace3}), we obtain

\begin{equation}
\begin{split}
&\Tr(\left.D_{\lat}^{(m)\,-1}\right\vert_{\cond{h^r}{h^i}}) = \int_0^{\infty}d\eta\,\rho_a(m,V,\eta)g(m,\eta) + \frac{a}{2}\frac{l}{V},\quad\Tr(\left.D_{\lat}^{(m)\,-1}\right\vert_{\cond{g}{h^i}}) = 
\int_0^{\infty}d\eta\,\rho_a(m,V,\eta)h^r(m,\eta) + \frac{a}{2}\frac{l}{V}.\\
\end{split}
\label{eq:qtrace1}
\end{equation}

\noindent
Hence the trace of eq. (\ref{eq:qprop1}) becomes 

\begin{equation}
\begin{split}
&\Tr (D_{\lat}^{(m)\,-1}) = a\frac{l}{V}\, +\int_0^{\infty}d\eta\,\rho_a(m,V,\eta)(g(m,\eta) + h^r(m,\eta)).
\end{split}
\label{eq:qtrace2}
\end{equation}

\noindent
Now we take the thermodynamic, continuum and massless limit of both sides in eq. (\ref{eq:qtrace1}), keeping in mind the procedure of such limits given in section \ref{sec:gap}. 
For the first equation in (\ref{eq:qtrace1}) we have 

\begin{equation}
\begin{split}
&\lc \Tr(\left.D_{\lat}^{(m)\,-1}\right\vert_{\cond{h^r}{h^i}}) =\lc 	\int_0^{\infty}d\eta\,\rho_a(m,V,\eta)g(m,\eta) \\
&=\lim_{m\To 0}\lim_{a\To 0}\int_0^{\infty}d\eta\,\rho_a(m,\eta)g(m,\eta) = \lim_{m\To 0}\int_0^{\infty}d\eta\,\rho_0(m,\eta)\tilde{g}(m,\eta) =\frac{\pi}{2}\rho_0(0,0),
\end{split}
\label{eq:glim}
\end{equation}

\noindent
where in the first line we have canceled the term $a\,l/V$, since in the limits $a\To 0$ and $V\To\infty$, it goes to zero, supposing that the multiplicity $l$ doesn't grow faster than $V/a$ on those limits, otherwise we would have an accumulation of eigenvalues at infinity. 
We also point out that such terms are always there for every configuration on which we want to estimate such trace, whatever $m$, $a$ and $V$. 
In the second line of (\ref{eq:glim}) we have passed the limit $V\To\infty$ inside the integral and in the third line we defined $\tilde{g}(m,\eta) = m/(m^2 + \eta^2)$. 
In the last step, we used that $\tilde{g}(m,\eta)$ is a Cauchy function which becomes a $\delta$-function in the limit $m\To 0$.

Regarding the second equation in (\ref{eq:qtrace1}), we can do similar considerations obtaining that 

\begin{equation}
\begin{split}
&\lc \Tr(\left.D_{\lat}^{(m)\,-1}\right\vert_{\cond{g}{h^i}}) =\lc 	\int_0^{\infty}d\eta\,\rho_a(m,V,\eta)h^r(m,\eta) \\
&=\lc \frac{\omega\, a}{k^2}\int_0^{\infty}d\eta\,\rho_a(k\,m',V,\eta)\frac{\eta^2}{\eta^2 + m'^2}=0\cdot\lc\int_0^{\infty}d\eta\,\rho_a(k\,m',V,\eta)\frac{\eta^2}{\eta^2 + m'^2} = 0,
\end{split}
\label{eq:hrlim}
\end{equation}

\noindent
where we used the expression of $h^r(m,\eta)$ in (\ref{eq:ghh}) and the introduction of the variable $m' = m/k$. 
In the third line we have split the two limits and we used that the integral $\int_0^{\infty}d\eta\,\rho_a(k\,m',V,\eta)\frac{\eta^2}{\eta^2 + m'^2}$ is finite. 
In fact $\vert\int_0^{\infty}d\eta\,\rho_a(k\,m',V,\eta)\frac{\eta^2}{\eta^2 + m'^2}\vert \leq \int_0^{\infty}d\eta\,\rho_a(k\,m',V,\eta) \leq 1$. 
Hence the last limit is finite and the product is zero. 

At this point we can take the limit $\lc$ on both sides of eq. (\ref{eq:qpropdis1}) and use the two results in (\ref{eq:glim}) and (\ref{eq:hrlim}), 
where the procedure introduced in section \ref{sec:gap}, for defining $\lc$, has been used, hence we have

\begin{equation}
\begin{split}
&\lc\left\vert \left.D_{\lat}^{(m)\,-1}(x,y)_{\alpha a, \beta b}\right\vert_{\cond{h^r}{h^i}} \right\vert\leq \frac{\pi}{2}\rho_0(0,0),\qquad\lc \left\vert \left.D_{\lat}^{(m)\,-1}(x,y)_{\alpha a, \beta b}\right\vert_{\cond{g}{h^i}} \right\vert = 0,
\end{split}
\label{eq:qpropdis2}
\end{equation}

\noindent
valid for all $x,y,\alpha,\beta,a,b$.
Those inequalities relates the quark propagator with the eigenvalue distribution in the sector of small $\eta$, when the limits $ \lim_{m\To0}\lim_{a\To0}\lim_{V\To\infty}$ are taken in such order.

\section{Left and right components of the quark propagator}\label{app:suLR}

In this appendix we prove eq. (\ref{eq:qpropRL}). 

At first we use that $\psi = \psi_L + \psi_R$ and $\bar{\psi} = \bar{\psi}_L + \bar{\psi}_R$, then we can decompose $(D_{\gap}^{-1})_{ij}$ as, 

\begin{equation}
\begin{split}
(D_{\gap}^{-1})_{ij} &= \lc \langle (\psi_{L,i} + \psi_{R,i})
(\bar{\psi}_{L,j} + \bar{\psi}_{R,j})\rangle\\ 
&= 
\lc \langle \psi_{L,i}\bar{\psi}_{L,j}\rangle + 
\lc \langle \psi_{L,i}\bar{\psi}_{R,j}\rangle+\lc \langle \psi_{R,i}\bar{\psi}_{L,j}\rangle+
\lc \langle \psi_{R,i}\bar{\psi}_{R,j}\rangle.
\end{split}
\label{eq:rlgap0}
\end{equation}

However $\lc \langle \psi_{R,i}\bar{\psi}_{L,j}\rangle$ and $\lc \langle \psi_{L,i}\bar{\psi}_{R,j}\rangle$ are null. 
Indeed, we can expand $\lc \langle \psi_{R,i}\bar{\psi}_{L,j}\rangle$,

\begin{equation}
\begin{split}
&\lc \langle \psi_{R,i}\bar{\psi}_{L,j}\rangle = 
\lc \left(P_R \langle \psi_{i}\bar{\psi}_{j}\rangle P_L\right) =
P_R\left(\lc \langle \psi_{i}\bar{\psi}_{j}\rangle\right) P_L = P_R\, \delta_{ij}D_{\gap}^{-1}P_L\\
&=\frac{\delta_{ij}}{4}(\mathds{1}+\gamma_5)D_{\gap}^{-1}(\mathds{1}+\gamma_5)=\frac{\delta_{ij}}{4}(
D_{\gap}^{-1}+
\gamma_5 D_{\gap}^{-1}+
D_{\gap}^{-1} \gamma_5
+\gamma_5 D_{\gap}^{-1} \gamma_5
),
\end{split}
\label{eq:rlgap}
\end{equation}

\noindent
where in the first line we used the definition of $\psi_R$ and $\bar{\psi}_L$, in the second line we have exchanged the operators $P_{L/R}$ with the limits $\lc$, and used that $(D_{\gap}^{-1})_{ij} = \delta_{ij}D_{\gap}^{-1}$, as described in section \ref{sec:axial}. 
Now using eq. (\ref{eq:gamma5anticomm}) we have that the last line of (\ref{eq:rlgap}) is zero, since $\gamma_5 D_{\gap}^{-1} \gamma_5 = -D_{\gap}^{-1}$, hence 
$
\lc \langle \psi_{R,i}\bar{\psi}_{L,j}\rangle = 0\quad\mbox{and}\quad
\lc \langle \psi_{L,i}\bar{\psi}_{R,j}\rangle = 0 $, 
where the second equation is obtained as in (\ref{eq:rlgap}) just exchanging $R\leftrightarrow L$. 
Finally using the definitions right after eq. (\ref{eq:qpropRL}), we have that eq. (\ref{eq:rlgap0}) can be rewritten as in eq. (\ref{eq:qpropRL}).

\section{Gap and correlators}\label{app:corr}

Here we want show how from eq. (\ref{eq:corr1}) we can arrive to the eq. (\ref{eq:corr2}), under the limits $\lc$ and when there is a gap in the eigenvalue distribution of the Dirac operator. 

At first we can take the fermionic average on both side of (\ref{eq:corr1}),

\begin{equation}
\langle\mathcal{C}(\psi,\bar{\psi})\rangle_F = 
\sum_{k}\sumcorrmin\langle\acorrkmin \prod_{l=1}^{k}\psi(x_l)_{I_l}\bar{\psi}(x_{k+l})_{I_{k+l}}\rangle_F,
\label{eq:corr1app}
\end{equation}

\noindent
where we passed the average as $\langle \cdot\rangle_F$ inside the sum. 
Now using the Wick theorem we have 

\begin{equation}
\begin{split}
&\langle\prod_{l=1}^{k}\psi(x_l)_{I_l}\bar{\psi}(x_{k+l})_{I_{k+l}}\rangle_F=\sum_p s(p)\prod_{l=1}^k\delta_{p(i_l)i_{k+l}}
\langle\psi(x_{p(l)})_{p(I_l)}\bar{\psi}(x_{k+l})_{I_{k+l}}\rangle_F\\
&=\sum_p s(p)\prod_{l=1}^k\delta_{p(i_l)i_{k+l}}
D^{(m)\,-1}_{\lat}(x_{p(l)},x_{k+l})_{p(\alpha_{l})p(a_l),\alpha_{k+l},a_{k+l}}
\end{split}
\end{equation}

\noindent
where, as explained in section \ref{sec:corr}, we have written explicitly the multi-index $I_{l} = \{i_l,\alpha_{l},a_{l}\}$, which indicates flavor, Dirac and color indices respectively. 
The label $p(l)$ (as well as $p(\alpha_{l})$ and $p(a)$) is obtained after the application of $p$ transpositions of $l$ (respectively $\alpha$ and $a$), and we are summing over all possible transpositions $p$; $s(p)$ is the sign which each permutation gives, coming from the exchange of $I_l$ and $p(I_l)$.

Now if we take the limits $\lc$ on both sides of eq. (\ref{eq:corr1app}), then we get

\begin{equation}
\begin{split}
&\lc\,\langle\mathcal{C}(\psi,\bar{\psi})\rangle_F= 
\sum_{k}\sumcorrmin\sum_p s(p)\, \acorrkbarmin\prod_{l=1}^k \delta_{p(i_l)i_{k+l}} (\lc D^{(m)\,-1}_{\lat}(x_{p(l)},x_{k+l})_{p(\alpha_{l})p(a_l),\alpha_{k+l} a_{k+l}})
\end{split}
\end{equation}

\noindent
where we defined $ \bar{\Gamma} = \lc \Gamma$. 
Then, under hypothesis of a gap in the Dirac spectrum, we have that $ D_{\gap}^{-1}\underset{\gap}{=}\lc D_{\lat}^{(m)\,-1}$, hence we obtain eq. (\ref{eq:corr2}).

\section{Chiralspin symmetry and quark propagator}\label{app:su2cs}

Here we want to show how from eq. (\ref{eq:qprop5}) we get the condition in eq. (\ref{eq:cond1}) and consequently the eq. (\ref{eq:cond2}), just imposing the invariance of $D^{-1}_{\gap}$ under the $SU(2)_{CS}$ transformations, which are given in eq. (\ref{eq:su2cs1}). 

At first we need to observe that $SU(2)_{CS}$ has three $U(1)$ subgroups, one for each generator. 
We can see, in fact, that for $\alpha_a = \{0,0,\alpha_3\}$, the transformations (\ref{eq:su2cs1}) are the same of eq. (\ref{eq:u1atrans1}), just in the point $x= (x_4,\bm{0})$, hence the axial group is a subgroup of the chiralspin group, namely $U(1)_A \subset SU(2)_{CS}$. 
Another $U(1)$ subgroup is generated by $\gamma_4$ which is defined, in $x= (x_4,\bm{0})$, as 

\begin{equation}
\begin{split}
U(1)_{4}:\quad&\psi(x_4)_i\To\exp(\I\alpha_1 \gamma_4)\,\psi(x_4)_i,\bar{\psi}(x_4)_i\To\bar{\psi}(x_4)_i\,\exp(-\I\alpha_1 \gamma_4),
\end{split}
\label{eq:su2cs2}
\end{equation}

\noindent
where we have just set $\alpha_a = (\alpha_1,0,0)$ in eq. (\ref{eq:su2cs1}). 
The other $U(1)$ subgroup is generated by $\I\gamma_5\gamma_4$, hence

\begin{equation}
\begin{split}
U(1)_{4A}:\quad&\psi(x_4)_i\To\exp(\I\alpha_2 (\I\gamma_5\gamma_4))\,\psi(x_4)_i,\bar{\psi}(x_4)_i\To\bar{\psi}(x_4)_i\,\exp(\I\alpha_2 (\I\gamma_5\gamma_4)),
\end{split}
\label{eq:su2cs3}
\end{equation}

\noindent
where we have just set $\alpha_a = (0,\alpha_2,0)$ in eq. (\ref{eq:su2cs1}). 

Since they are three distinct $U(1)$ subgroups, in order to impose the invariance of $D^{-1}_{\gap}(x_4,y_4)$ under $SU(2)_{CS}$, is sufficient to impose the invariance under these three $U(1)$ subgroup transformations. 
However, from section \ref{sec:axial}, we have shown that $D^{-1}_{\gap}$ is already invariant under $U(1)_A$, hence $D^{-1}_{\gap}(x_4,y_4)$ will be also invariant. 
Therefore we just need to consider the other two subgroups.
Regarding $U(1)_{4}$, we need to plug the transformations (\ref{eq:su2cs2}), inside eq. (\ref{eq:qprop5}), then $D^{-1}_{\gap}(x_4,y_4)$, transforms as 

\begin{equation}
\begin{split}
&D^{-1}_{\gap}(x_4,y_4)\To D^{-1}_{\gap}(x_4,y_4)^{U(1)_{4}} = \cos^2(\alpha_1)\, D^{-1}_{\gap}(x_4,y_4)\\
&+\I \sin(\alpha_1) \cos(\alpha_1)\,(\gamma_4 D^{-1}_{\gap}(x_4,y_4) -D^{-1}_{\gap}(x_4,y_4)\gamma_4)+\sin^2(\alpha_1)\, \gamma_4D^{-1}_{\gap}(x_4,y_4)\gamma_4.
\end{split}
\label{eq:transcs1}
\end{equation}

\noindent
Instead if we plug (\ref{eq:su2cs3}) inside eq. (\ref{eq:qprop5}) we get

\begin{equation}
\begin{split}
&D^{-1}_{\gap}(x_4,y_4)\To D^{-1}_{\gap}(x_4,y_4)^{U(1)_{4A}} = \cos^2(\alpha_2)\, D^{-1}_{\gap}(x_4,y_4)\\
&+\I \sin(\alpha_2) \cos(\alpha_2)\,(\I\gamma_5\gamma_4 D^{-1}_{\gap}(x_4,y_4) +D^{-1}_{\gap}(x_4,y_4)\I\gamma_5\gamma_4)-\sin^2(\alpha_2)\, \gamma_5\gamma_4 D^{-1}_{\gap}(x_4,y_4)\gamma_5\gamma_4.
\end{split}
\label{eq:transcs2}
\end{equation}

\noindent
From (\ref{eq:transcs1}) and (\ref{eq:transcs2}), we can see that, differently from $SU(n_F)_L\times SU(n_F)_R$ and $U(1)_A$, $D^{-1}_{\gap}(x_4,y_4)$ is not invariant under $SU(2)_{CS}$.
However it would be invariant if and only if $\gamma_4 D^{-1}_{\gap}(x_4,y_4)\gamma_4 = D^{-1}_{\gap}(x_4,y_4)$. 
This condition can be rewritten in terms of the left and right components of the eigenvectors $v_n^{(\pm)}$, given in eq. (\ref{eq:eigvect}). 
For doing this, we need to consider the expression of $D^{-1}_{\gap}(x_4,y_4)$ in eq. (\ref{eq:qprop4}), and passing $\gamma_4$ inside $\lc$ and the sum over $\eta_n$, then we get that the condition $\gamma_4 D^{-1}_{\gap}(x_4,y_4)\gamma_4 = D^{-1}_{\gap}(x_4,y_4)$ is equivalent to

\begin{equation}
\begin{split}
&\sum_{\eta_n>\Lambda} \I h^i(m,\eta_n)\gamma_4\left(\begin{matrix}
0 & L_n(x_4) R_n^{\dagger}(y_4)\\
R_n(x_4) L_n^{\dagger}(y_4) & 0 
\end{matrix}\right)\gamma_4\quad \underset{\CS}{=} \sum_{\eta_n>\Lambda} \I h^i(m,\eta_n)\left(\begin{matrix}
0 & L_n(x_4) R_n^{\dagger}(y_4)\\
R_n(x_4) L_n^{\dagger}(y_4) & 0 
\end{matrix}\right)\\
&\Rightarrow
\sum_{\eta_n>\Lambda} \I h^i(m,\eta_n)
(L_n(x_4) R_n^{\dagger}(y_4)-R_n(x_4)L_n^{\dagger}(y_4)) \underset{\CS}{=} 0,
\end{split}
\label{eq:condcs2}
\end{equation}

\noindent
and assuming 
the last line of (\ref{eq:condcs2}) is zero if each term inside the sum is zero, then we get that $L_n(x_4) R_n^{\dagger}(y_4)-R_n(x_4)L_n^{\dagger}(y_4) = 0$, which is the condition in eq. (\ref{eq:cond2}).

Finally, calling $D^{-1}_{\gap,CS}(x_4,y_4)$ the quark propagator $D^{-1}_{\gap}(x_4,y_4)$ satisfying the condition in eq. (\ref{eq:cond1}), then its invariance over the three $U(1)$ subgroups of $SU(2)_{CS}$, which is generated by the matrices: $\{\gamma_4,\I\gamma_5\gamma_4,-\gamma_5\}$, corresponds to an invariance of $D^{-1}_{\gap,CS}(x_4,y_4)$ over the full $SU(2)_{CS}$ group. 


\bibliographystyle{unsrtnat}
\bibliography{u1a_v1}

\end{document}